\documentclass[prl,twocolumn,showpacs,floatfix,amsmath,amssymb,floatfix]{revtex4}

\usepackage{epsfig,epsf,graphics,psfrag}
\usepackage{times}
\usepackage{float}
\usepackage{color}
\usepackage{amsfonts,amssymb,stmaryrd,latexsym,amsmath}
\usepackage{multirow}
\usepackage{array} 
\usepackage{hhline}
\usepackage{booktabs}
\usepackage{enumerate}
\usepackage{slashed}
\usepackage{subfigure}
\usepackage[hyperindex=true]{hyperref}

\begin{document}

\bibliographystyle{unsrt}

\title{Generating a resonance-like structure  in the reaction {\boldmath$B_c\to B_s \pi\pi$}}

\author{Xiao-Hai Liu$^a$\footnote{xiaohai.liu@fz-juelich.de}}
\author{Ulf-G. Mei{\ss}ner$^{a,b}$\footnote{meissner@hiskp.uni-bonn.de} }

\affiliation{ $^a$Forschungszentrum J\"ulich, Institute for Advanced Simulation, Institut f\"ur Kernphysik and J\"ulich Center for Hadron Physics, D-52425 J\"ulich, Germany \\
$^b$ Helmholtz-Institut f\"ur Strahlen- und Kernphysik and Bethe Center for
Theoretical Physics,
Universit\"at Bonn, D-53115 Bonn, Germany	}

\date{\today}

\begin{abstract}

We investigate the process $B_c^+\to B_s^0\pi^+\pi^0$ via  $B\bar{K}^*$ rescattering. The kinematic conditions for  triangle singularities are perfectly satisfied in the rescattering diagrams. A resonance-like structure around the $B\bar{K}$ threshold, which we denote as $X(5777)$, is predicted to be present in the invariant mass distribution of $B_s^0 \pi^+$. Because the relative weak $B\bar{K}$ $(I=1)$ interaction does not support the existence of a dynamically generated hadronic molecule, the $X(5777)$ can be identified as a pure kinematical effect due to the triangle singularity. Its observation may help  to establish a non-resonance interpretation for some $XYZ$ particles.




\end{abstract}

\maketitle

\noindent\emph{$\;$Introduction}.\,---\,%
 Hadron spectroscopy, in particular due to  the appearance of the  so-called exotic states, is experiencing a renaissance in recent years. Since 2003, dozens of resonance-like structures have been observed by many experimental collaborations in various reactions. These structures are usually denoted as $XYZ$ particles, because most of them do not fit into the conventional quark model (QM), which has been very successful in describing the low-lying hadrons. For instance, the observed masses of the $X(3872)$ and  the $D_{s0}^*(2317)$ are much smaller than the expected values of the conventional QM states $\chi_{c1}(2 ^3P_1)$ and $D_{s0}^* (1 ^3P_0)$, respectively. Some of these states definitely cannot be conventional $q\bar{q}$-mesons or $qqq$-baryons, such as the charged $Z_c^\pm$/$Z_b^\pm$ states observed in $J/\psi\pi^\pm$/$\Upsilon(nS)\pi^\pm$ invariant mass distributions, the $P_c(4380)$ and $P_c(4450)$ observed in $J/\psi p$ distributions, and so on. These experimental observations have also inspired a flurry of theoretical investigations trying to understand their intrinsic structures. We refer to 
Refs.~\cite{Chen:2016spr,Brambilla:2010cs,Olsen:2014qna,Brambilla:2014jmp,Chen:2016qju,Esposito:2016noz}
for some recent reviews about the study of exotic hadrons.
 Among the popular theoretical interpretations about exotic hadrons, the multi-quark (tetraquark, pentaquark, etc.) interpretation usually tends to imply the existence of a large number of degenerate states. In contrast, the observed spectrum in experiments appears to be very sparse, which is a challenge for this interpretation. An intriguing characteristic of the $XYZ$ states is that many of them are located around two-meson (or one meson and one baryon) thresholds. For example, the masses of the $D_{s0}^*(2317)$, $X(3872)$, $Y(4260)$, $Z_b(10610)$ and $Z_{b}(10650)$ are very close to the threshold of $DK$, $D\bar{D}^*$, $D_1\bar{D}$, $B\bar{B}^*$ and $B^*\bar{B}^*$, respectively. This phenomenon can be considered an evidence for regarding some $XYZ$ states as hadronic molecules -- bound systems of two hadrons analogous to conventional nuclei. The deuteron, which is composed of a proton and a neutron, is the one of the few well established hadronic molecule up to now. With proper interactions, the existence of molecular states composed of other hadrons is also expected. A prime example is the $\Lambda(1405)$, which was predicted as a $\bar KN$ molecule long before the QM. In many cases, however, the detailed multi-hadron dynamics is not so well understood. For a recent review on hadronic molecules, see Ref.~\cite{RMP}.

 Concerning the underlying structures of those $XYZ$ states, besides genuine resonances interpretations mentioned above, some non-resonance interpretations which connect the kinematic singularities of the rescattering amplitudes with the resonance-like peaks were also proposed in literatures, such as the cusp model~\cite{Chen:2011pv,Bugg:2011jr,Swanson:2014tra},  or the triangle singularity (TS) mechanism. The TS mechanism  was first noticed in 1960s~\cite{Aitchison:1969tq,Coleman:1965xm,Bronzan:1964zz,Schmid:1967ojm}.  Unfortunately, most of the proposed reactions at that time were lacking  experimental data. It was rediscovered  in recent years and used to interpret some exotic phenomena, such as the large isospin violation in $\eta(1405)\to 3\pi$, the production of the axial-vector state $a_1(1420)$, the production of the $Z_c^\pm(3900)$ and so on\cite{Wu:2011yx,Wang:2013cya,Guo:2014iya,Ketzer:2015tqa,Szczepaniak:2015eza,Guo:2015umn,Liu:2015taa,Liu:2015fea,Achasov:2015uua,Guo:2016bkl,Roca:2017bvy}.
 It is shown that sometimes it is not necessary to introduce a genuine resonance to describe a resonance-like peak, because the TSs of the rescattering amplitudes could generate bumps in the corresponding invariant mass distributions. Before claiming that a resonance-like peak corresponds to a genuine particle, it is also necessary to exclude or confirm the possibility of this non-resonance interpretation. As for the cusp model, it should be mentioned that in Ref.~\cite{Guo:2014iya} it was shown that the kinematic threshold cusp cannot produce a narrow peak in the invariant
 mass distribution of the elastic channel in contrast with a genuine $S$-matrix pole.  
 
 The position of  the TS peak usually stays in the vicinity of the threshold of the scattering particles. From this point of view, the TS mechanism is similar to the hadronic molecule interpretation, and it also implies that the genuine dynamic pole may mix with the TS peak. This brings some ambiguities to our understanding about the nature of some resonance-like peaks observed in experiments. One way to distinguish TS peaks from genuine resonances is finding some ``clean" processes.  Since the pole position of a genuine state should not depend on a specific process, while the TS peak is rather sensitive to the kinematic conditions, one would expect that a genuine state should still appear in the processes where kinematic conditions for the TS are not fulfilled, but the TS peak should disappear. Vice versa, if one observes a resonance-like peak in a process where the genuine state does not contribute but the TS kinematic conditions can be fulfilled, it will also help to establish the TS mechanism. In this paper, we focus on a process through which the TS mechanism could be confirmed in experiments.
 
\begin{figure}[htbp]
	\centering
	\includegraphics[width=0.7\hsize]{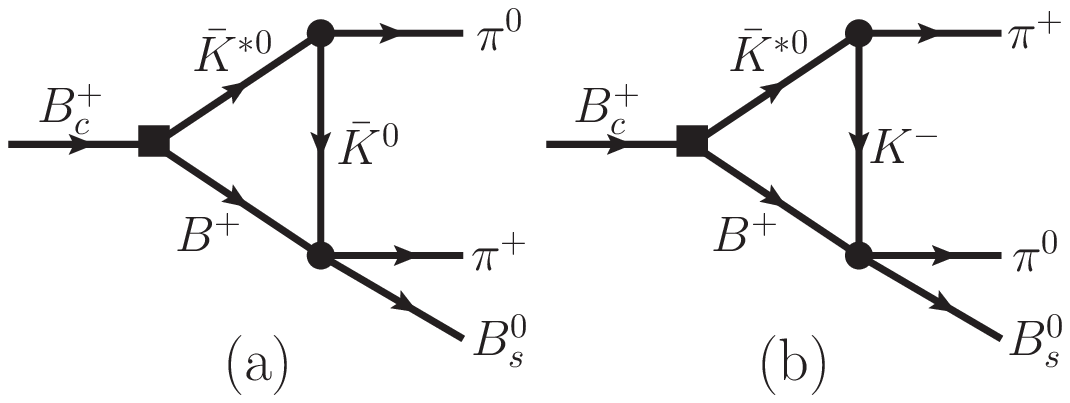}
	\caption{$B_c^+\to B_s^0  \pi^+\pi^0$ via the triangle rescattering diagrams.}\label{Triangle-Diagram}
\end{figure}

\noindent\emph{$\;$TS Mechanism}.\,---\,
 For the triangle Feynman diagrams describing rescattering processes, such as those illustrated in Fig.~\ref{Triangle-Diagram}, there are two kinds of intriguing singularities which may appear in the rescattering amplitudes. When only two of the three intermediate states are on-shell, the singularity at threshold is a finite square-root branch point, which corresponds to a cusp effect. In some special kinematical configurations, all of the three intermediate states can be on-shell simultaneously, which corresponds to the leading Landau singularity of the triangle diagram. This leading Landau singularity is usually called the TS, which may result in a narrow peak in the corresponding spectrum. 
 
 For the decay process $B_c^+\to B_s^0  \pi^+\pi^0$ via the $\bar{K}^*B \bar{K}$-loop in Fig.~\ref{Triangle-Diagram}(a), we define the invariants $s_1\equiv p_{B_c^+}^2=m_{B_c^+}^2$, and $s_2\equiv (p_{B_s^0}+p_{\pi^+})^2=M_{B_s^0\pi^+}^2$. The position of the TS in the $s_1$ or $s_2$ complex plane of the scattering amplitude $\mathcal{A}(s_1,s_2)$ can be determined by solving the so-called Landau equation~\cite{Landau:1959fi,Coleman:1965xm}.
 Assuming we do not know the physical mass $m_{B_c^+}$, when  $\sqrt{s_1}$ increases from the $B\bar{K}^*$ threshold $6.175$~GeV to $6.297$~GeV, the TS in $\sqrt{s_2}$ moves from $5.849$~GeV to the $B\bar{K}$ threshold 
at $5.777$~GeV. Vice versa, when $\sqrt{s_2}$ increases from $5.777$ to $5.849$~GeV, the TS in $\sqrt{s_1}$ moves from $6.297$ to $6.175$ GeV. These are the kinematical regions where the TS can be present in the physical rescattering amplitude. It is interesting to note that the mass of $B_c^+\sim 6.276$~GeV just falls into the TS kinematical region. 

Taking Fig.~\ref{Triangle-Diagram}(a) as an example, the physical picture concerning the TS mechanism can be understood like this: The initial particle $B_c^+$ first decays into $B^+$ and $\bar{K}^{*0}$, then the particle $\bar{K}^0$ emitted from $\bar{K}^{*0}$ catches up with the $B^+$, and finally $B^+\bar{K}^0$  scatters into $B_s^0 \pi^+$. This implies that the rescattering diagram can be interpreted as a classical process in space-time with the presence of TS, and the TS will be located on the physical boundary of the rescattering amplitude~\cite{Coleman:1965xm}.

\noindent\emph{$\;$Rescattering Amplitude}.\,---\,%
The $B_c^+$ meson, lying below the $BD$ threshold, can only decay via the weak interactions, and about $70\%$ of its width is due to $c$ quark decay with the $b$ quark as  spectator~\cite{Brambilla:2004wf}. The decay $B_c^+\to B^+ \bar{K}^{*0} $, as a Cabibbo-favored process, is expected to be one of the dominant nonleptonic decay modes of the $B_c^+$ \cite{Du:1988ws,Sun:2015exa}. There is no direct measurement on this channel at present. Its branching ratio is usually predicted to be larger than $10^{-3}$, which implies the rescattering processes in Fig.~\ref{Triangle-Diagram} may play a role in $B_c^+\to B_s^0\pi^+\pi^0$. Following Refs.~\cite{Du:1988ws,Sun:2015exa}, by means of the factorization approach, the decay amplitude can be expressed as
\begin{eqnarray}\label{BctoBKstar}
	 \mathcal{A}(B_c^+\to B^+ \bar{K}^{*0}) &=& \sqrt{2} G_F F_1^{B_c \to B_u} f_{{K}^*} m_{\bar{K}^*} \nonumber \\ 
	&\times& (p_{B_c^+}\cdot \epsilon^*_{\bar{K}^*}) V_{ud}^{} V_{cs}^* a_2,
\end{eqnarray}
where $G_F$ is the Fermi coupling constant, $F_1^{B_c \to B_u}$ is the $B_c^+\to B^+$ transition form factor, $f_{{K}^*}$ is the decay constant of the $K^*$, $V_{ud}^{}$,$V_{cs}^*$ are the CKM matrix elements, and $a_2$ is a combination of Wilson coefficients. For $B_c^+\to B^+ \bar{K}^{*0} $, the velocity of the recoiling $B^+$ is very low in the rest-frame of the $B_c^+$, and the wave functions of $B_c^+$ and $B^+$ overlap  strongly. The form factor $F_1^{B_c \to B_u}$ is then expected to be close to unity \cite{Du:1988ws,Sun:2015exa,Wirbel:1985ji}. In our numerical calculation, we take  $F_1^{B_c \to B_u}=1$ as an approximation. The  decay constant $f_{K^*}$ and coefficient $a_2$ are fixed to be $220\ \mbox{MeV}$ and $-0.4$, respectively \cite{Sun:2015exa}. Concerning the other parameters in Eq.~(\ref{BctoBKstar}), we input the standard Particle Data Group values ~\cite{Olive:2016xmw}. 

For $\bar{K}^*\to \bar{K}\pi$, the amplitudes take the form
\begin{eqnarray}
	\mathcal{A}(\bar{K}^{*0}\to \bar{K}^0\pi^0)&=& 2G_V p_{\pi^0} \cdot \epsilon_{\bar{K}^*}, \\
	\mathcal{A}(\bar{K}^{*0}\to K^-\pi^+)&=& -2\sqrt{2} G_V p_{\pi^+} \cdot \epsilon_{\bar{K}^*},
\end{eqnarray}
where the coupling constant $G_V$ can be determined by the decay width of the $\bar{K}^*$.

There have been many theoretical studies about the pesudo-Nambu-Goldstone-bosons ($\pi$, $K$, etc.) scattering off the heavy-light mesons ($D^{(*)}$, $B^{(*)}$, etc.). 
By means of  lattice QCD (LQCD) simulations and chiral extrapolation, in Ref.~\cite{Liu:2012zya} the $S$-wave scattering length of the isoscalar $DK$ channel $a_{DK}^{I=0}$ is predicted to be $-0.86\pm 0.03\,$fm at the physical pion mass. Employing both $\bar{s}c$ and $DK$ interpolating fields, in Ref.~\cite{Mohler:2013rwa} the authors performed a direct lattice simulation and obtain the $DK$ scattering length $a_{DK}^{I=0}= -1.33(20)\,$fm, which qualitatively agrees with the result of Ref.~\cite{Liu:2012zya}. The large negative scattering length $a_{DK}^{I=0}$ means the $DK$ ($I=0$) interaction is strong, and indicates the presence of an isoscalar state below threshold. It is generally supposed that the $D_{s0}^{*}(2317)$/$D_{s1}(2460)$ is the hadronic molecule dynamically generated by the strong $DK$/$D^*K$ ($I=0$) interaction in the coupled-channels dynamics \cite{Guo:2006fu,Guo:2009ct,Liu:2009uz,Kolomeitsev:2003ac,Altenbuchinger:2013vwa,Guo:2015dha,Liu:2012zya,Mohler:2013rwa,Chen:2016spr}. On the other hand, the scattering length of the isospin-1 $DK$ channel $a_{DK}^{I=1}$ is predicted to be $0.07\pm 0.03 + i(0.17^{+0.02}_{-0.01})\,$fm in Ref.~\cite{Liu:2012zya}, which is much smaller than $a_{DK}^{I=0}$ and implies the $DK$ ($I=1$) interaction is weak. According to the heavy quark spin and flavor symmetry, the above results can be easily extended to the ${B}^{(*)} \bar{K}$ cases. The bottom-quark counterparts of $D_{s0}^{*}(2317)$ and $D_{s1}(2460)$ are ${B}_{s0}^*$ and ${B}_{s1}$, which are supposed to be the ${B}\bar{K}$ and ${B}^*\bar{K}$ molecular states, respectively. But these two states have not been observed in experiments yet. The predicted masses of ${B}_{s0}^*$/${B}_{s1}$ are usually tens of MeV below the  ${B}\bar{K}$/${B}^*\bar{K}$ threshold. Being similar to the $DK$ ($I=1$) interaction, the $B\bar{K}$ ($I=1$) interaction is also generally supposed to be weak. Within the framework of an unitary chiral effective field theory, the $S$-wave scattering length of isovector $B\bar{K}$ channel $a_{B\bar{K}}^{I=1}$ is predicted to be $0.02-0.23i\,$fm ~\cite{Altenbuchinger:2013vwa,Lu:2016kxm}. The relative weak interactions in the $B\bar{K}$-$B_s\pi$ coupled-channels do not support the presence of an isovector hadronic molecule around $B\bar{K}$ threshold.

In 2016, the D0 collaboration reported the observation of a narrow structure $X(5568)$ in the $B_s^0\pi^\pm$ invariant mass spectrum~\cite{D0:2016mwd}. The mass and width are measured to be $M_X=5567.8\pm 2.9^{+2.9}_{-1.9}$ MeV and $\Gamma_X=21.9\pm 6.4^{+5.0}_{-2.5}$ MeV, respectively. The quark components of the decaying final state $B_s^0 \pi^\pm$ are $su\bar b \bar d$ (or $sd\bar b \bar u$), which requires $X(5568)$ should be a structure with four different valence quarks. Considering its mass and quark contents, some theorists suppose it could be an isovector hadronic molecule composed of $B\bar{K}$ \cite{Albaladejo:2016eps}. Using a chiral unitary approach,  the authors reproduce the reported spectrum of D0 collaboration. However, the authors of Ref.~\cite{Albaladejo:2016eps} also point out to reproduce the spectrum an  ``unnatural'' cutoff $\Lambda \simeq 2.8\ \mbox{GeV}$ is adopted in the $T$-matrix regularization, which is much larger than the ``natural'' value $\Lambda \simeq 1\ \mbox{GeV}$. Furthermore, in Ref.~\cite{Albaladejo:2016eps}, only the leading order (LO) potential was adopted, but in Ref.~\cite{Altenbuchinger:2013vwa} it was shown that the LO potential cannot describe the  LQCD scattering lengths of Ref.~\cite{Liu:2012zya}. Employing the covariant formulation of the unitary chiral perturbation theory (UChPT), the authors found no bound state or resonant state via a direct searching on different Riemann sheets in Refs.~\cite{Altenbuchinger:2013vwa,Lu:2016kxm}, where the driving potentials up to next-to-leading order (NLO) are constructed.  
 In a recent experimental result reported by the LHCb collaboration \cite{Aaij:2016iev}, the existence of $X(5568)$ is not confirmed based on their $pp$ collision data, which makes the production mechanism and underlying structure of $X(5568)$ more puzzling. In fact, right after the observation by D0, the possible existence of 
this state was challengend on theoretical grounds, see Refs.~\cite{Burns:2016gvy,Guo:2016nhb}.
The reason of its appearance in the D0 and absence in LHCb and CMS experiments is discussed in 
Ref.~\cite{Yang:2016sws}.

What we are interested in this paper is not the $X(5568)$ but a predicted resonance-like peak 
denoted as $X^\pm (5777)$ located around the $B\bar{K}$ threshold in the $B_s\pi^\pm$ distributions. 
Because the existence of an isovector $B\bar{K}$ hadronic molecule is rather questionable, 
for the decay process $B_c^+\to B_s^0 \pi^+\pi^0$, if one finds a peak in the $B_s^0\pi^+$ invariant 
mass spectrum around $B\bar{K}$ threshold, it is quite reasonable to suppose that the peak is 
induced by the TS mechanism as illustrated in Fig.~\ref{Triangle-Diagram}.

For the vertex $B\bar{K}\to B_s\pi$ in Fig.~\ref{Triangle-Diagram}, we employ the amplitude which is unitarized according to the method of UChPT ~\cite{Oller:2000fj,Oller:2000ma,Oller:1997ng}.
We consider the $S$-wave $B\bar{K}$ and $ B_s\pi$ coupled-channel scattering. The unitary $T$-matrix is given by
\begin{eqnarray}
T=(1-VG)^{-1}V,
\end{eqnarray}
where $V$ stands for the $S$-wave projection of the driving potential, 
 and $G$ is a diagonal matrix composed of two-meson-scalar-loop functions \cite{Oller:2000fj,Oller:2000ma,Oller:1997ng}. We only focus on the $S$-wave scattering in this paper, because the higher partial wave contributions will be highly suppressed for the near-threshold scattering. In our numerical calculations, the NLO potential is used. For the pertinent low-energy-constants and subtraction constant, we adopt the values of Ref.~\cite{Altenbuchinger:2013vwa}, which are determined by fitting the recent LQCD result of Ref.~\cite{Liu:2012zya}. See Refs.~\cite{Guo:2009ct,Guo:2015dha,Liu:2012zya,Altenbuchinger:2013vwa,Oller:2000fj} for more details about the formulation of NLO potentials.

The rescattering amplitude of $B_c^+\to B_s^0\pi^+\pi^0$ via the $\bar{K}^{*0}(q_1) B^+(q_2) \bar{K}^0 (q_3)$-loop in Fig.~\ref{Triangle-Diagram} (a) is given by
\begin{eqnarray}\label{amplitude-loop}
&&\mathcal{A}_{B_c^+\to B_s^0\pi^+\pi^0}^{[ \bar{K}^{*0} B^+ \bar{K}^0 ]} = \frac{1}{i} \int \frac{d^4q_3}{(2\pi)^4} \frac{\mathcal{A}(B_c^+\to B^+ \bar{K}^{*0})  }{ (q_1^2-m_{\bar{K}^*}^2 +i m_{\bar{K}^*}\Gamma_{\bar{K}^*})  }  \nonumber \\
&&\times \frac{ \mathcal{A}(\bar{K}^{*0}\to \bar{K}^0\pi^0) \mathcal{A}(B^+\bar{K}^0\to B_s^0\pi^+) }{ (q_2^2-m_{B^+}^2) (q_3^2-m_{\bar{K}^0}^2) } \mathbb{F}(q_3^2),
\end{eqnarray}
where the sum over polarizations of intermediate state is implicit. The amplitude of Fig.~\ref{Triangle-Diagram}(b) is similar to that of Fig.~\ref{Triangle-Diagram}(a). As long as the TS kinematical conditions are satisfied, it implies that one of the intermediate state ($\bar{K}^*$ here) must be unstable. It is necessary to take into account the width effect of intermediate state. We therefore employ a Breit-Wigner (BW) type propagator in Eq.~(\ref{amplitude-loop}). The complex mass in the propagator will remove the TS from physical boundary by a small distance, and makes the physical scattering amplitude finite. Since the location of TS is not far from the physical boundary, the physical amplitude can still feel its influence.
In Eq.~(\ref{amplitude-loop}), we also introduce a monopole form factor $\mathbb{F}(q_3^2)=(m_{\bar{K}}^2-\Lambda^2 )/(q_3^2-\Lambda^2)$ 
to account for the off-shell effect and kill the ultraviolet divergence that appears in the loop integral. 
In the future, this has to be replaced by a better regularization procedure.

\begin{figure}[htbp]
	\centering
	\includegraphics[width=0.7\hsize]{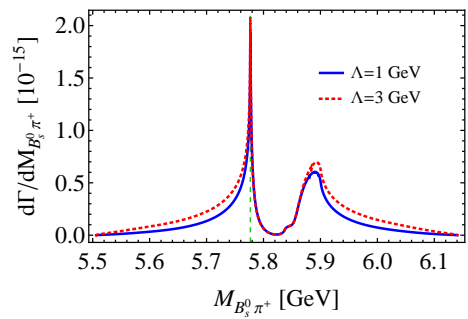}
	\caption{Invariant mass distributions of $B_s^0\pi^+$ via the triangle rescattering diagrams in Fig.~\ref{Triangle-Diagram}. The vertical dashed line indicates the $B^+\bar{K}^0$ threshold.}\label{signal}
\end{figure}

The numerical results of $B_s^0\pi^+$ distributions via the rescattering processes are displayed in Fig.~\ref{signal}, where the cutoff energy $\Lambda$ is taken to be 1~GeV or 3~GeV. It can be seen that the lineshape is not sensitive to the value of $\Lambda$. The two curves nearly coincide with each other, even though the cutoff energies are rather different. This is because the dominant contribution to the loop integral in Eq.~(\ref{amplitude-loop}) comes from the region where intermediate particles are (nearly) on-shell, i.e. when $q_3^2=m_{\bar{K}}^2$, $\mathbb{F}(q_3^2)$ gives 1.

A narrow peak around $5.777$ GeV  can be seen in Fig.~\ref{signal}. This resonance-like peak is what we 
call the $X(5777)$. As analyzed above, the $X(5777)$ discussed here is not a dynamically generated pole in the coupled-channel dynamics. Its presence is due to the TS kinematical conditions being fulfilled 
in the rescattering diagram. The bump around $5.9$~GeV in Fig.~\ref{signal} is due to reflection 
effects in the Dalitz plot and interference between Figs.\ref{Triangle-Diagram} (a) and (b).

\noindent\emph{$\;$Background Analysis}.\,---\,%
The rescattering processes in Fig.~\ref{Triangle-Diagram} is just one of the contributions to 
three-body decay $B_c^+\to B_s^0\pi^+\pi^0$. Since the TS peak can appear in these diagrams, 
we define them as the ``signal'' processes. But the dominant contribution to $B_c^+\to B_s^0\pi^+\pi^0$ 
is expected to be via the process $B_c^+\to B_s^0 \rho^+ \to B_s^0\pi^+\pi^0$. This is because 
compared to $B_c^+\to B_s^0 \rho^+ $, $B_c^+\to B^+ \bar{K}^{*0}$  is a color-suppressed process in 
the naive factorization approach. The branching ratio of $B_c^+\to B_s^0 \rho^+ $ is generally 
predicted to be larger than $1\%$ ~\cite{Sun:2015exa}, which is about one order of magnitude 
larger than that of $B_c^+\to B^+ \bar{K}^{*0}$. To study the ``signal" in the $B_s^0\pi^+$ 
distribution, it is also necessary to know the influence of possible backgrounds, 
especially the $B_c^+\to B_s^0 \rho^+ $. 

Using the factorization approach, the amplitude of $B_c^+\to B_s^0 \rho^+ $ can be written as
\begin{eqnarray}\label{BctoBsRho}
\mathcal{A}(B_c^+\to B_s^0 \rho^+) &=& \sqrt{2} G_F F_1^{B_c \to B_s} f_\rho m_\rho \nonumber \\ 
&\times& (p_{B_c^+}\cdot \epsilon^*_{\rho}) V_{ud}^{} V_{cs}^* a_1,
\end{eqnarray}
where we use $F_1^{B_c \to B_s}=1$, $f_\rho=216\ \mbox{MeV}$, $a_1=1.22$ in the numerical calculations \cite{Sun:2015exa}. The amplitude of $\rho^+\to\pi^+\pi^0$ reads $\mathcal{A}(\rho^+\to\pi^+\pi^0) = 4G_V p_{\pi^0} \cdot \epsilon_{\rho}$.
The complete amplitude of $B_c^+\to B_s^0\pi^+\pi^0$ is then given by
\begin{eqnarray}\label{total-amplitude}
	\mathcal{A}(B_c^+\to B_s^0\pi^+\pi^0) = 
  e^{i\theta}\mathcal{A}_\rho^{\mbox{tree}} +\mathcal{A}^{\mbox{loop}}\mathcal{F}(s_{\pi\pi}),
\end{eqnarray}
where $\mathcal{A}_\rho^{\mbox{tree}}$ is the amplitude of a tree diagram via intermediate $\rho$ meson decay, and the normal BW type propagator is adopted in $\mathcal{A}_\rho^{\mbox{tree}}$. The factor $e^{i\theta}$ stands for the relative phase between $\mathcal{A}(B_c^+\to B_s^0 \rho^+)$ and $\mathcal{A}(B_c^+\to B^+ \bar{K}^{*0})$, which is actually not fixed in the factorization approach. In the above equation, we also introduce a function $\mathcal{F}(s_{\pi\pi})$ to account for the strong $\pi\pi$ final-state-interaction ~\cite{Au:1986vs,Dai:2012pb,Wang:2013cya}, where $s_{\pi\pi}$ is $\pi^+\pi^0$ invariant mass squared. 
Due to the generalized Bose statistics, $\pi^+\pi^0$ can only stay in relative odd partial waves. For the lowest $P$-wave $\pi\pi$ scattering, the phase shift in the isospin-1 channel can be well reproduced by the intermediate $\rho$-meson exchange. The function $\mathcal{F}(s_{\pi\pi})$ can be further parametrized as $\mathcal{F}(s_{\pi\pi})=\alpha(s_{\pi\pi})/(s_{\pi\pi}-{m}_\rho^2+i m_\rho \Gamma_\rho)$. $\alpha(s_{\pi\pi})$ is a polynomial function of $s_{\pi\pi}$, which should be fixed according to the experimental data. But since we are going to make a prediction here, we approximately take  $\alpha(s_{\pi\pi})=s_{\pi\pi}-\overset{\circ}{m}_\rho^2$, where $\overset{\circ}{m}_\rho$ is the bare mass of $\rho$ meson without the effect of $\pi\pi$ meson loop. By reproducing the $P$-wave $\pi\pi$ scattering phase shift data,  $\overset{\circ}{m}_\rho$ is fixed to be $0.81$~GeV according to a vector-meson-dominance model employed in Ref.~\cite{Klingl:1996by}. This rather model-dependent scheme should eventually be replaced by taking a more improved spectral function, see e.g. Refs.~\cite{Daub:2012mu,Daub:2015xja,Chen:2015jgl,Chen:2016mjn} (and references therein).

\begin{figure}[htbp]
	\centering
	\includegraphics[width=0.7\hsize]{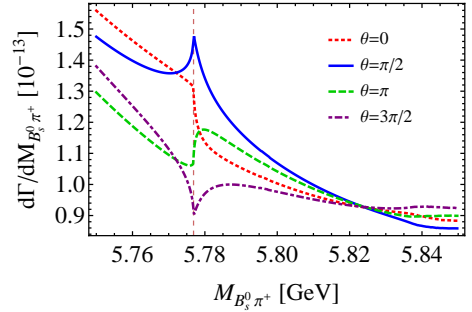}
	\caption{Simulated $B_s^0\pi^+$ distribution including both contributions of signal and background. The vertical dashed line indicates the $B^+\bar{K}^0$ threshold.}\label{Dalitz-plot}
	\vspace{-3mm}
\end{figure}

In terms of Eq.~(\ref{total-amplitude}), the simulated $B_s^0\pi^+$ distribution is displayed in Fig.~\ref{Dalitz-plot}, where the relative phase $\theta$ is taken to be $0$, $\pi/2$, $\pi$ and $3\pi/2$, respectively, corresponding to the different curves.
The cutoff energy $\Lambda$ is fixed to be $1\,$GeV in the simulation. The $B_s^0\pi^+$ distribution is dominated by the reflection of the $\rho$ signal in the Dalitz plot, but
all of the four curves in Fig.~\ref{Dalitz-plot} deviate significantly from the reflection around $5.777$~GeV. When $\theta=0\ (\pi)$, there is a sudden fall (rise) in the distributions. When $\theta=\pi/2\ (3\pi/2)$, there is a narrow peak (dip) in the distributions. The TS of the rescattering process generates  different structures due to different interferences.

Another background may come from the isospin-violation process $B_c^+\to B_{s0}^{*}\pi^+\to B_s^0\pi^0\pi^+$. But since the $B_{s0}^{*}$ peak in the $B_s^0\pi^0$ distribution may not have a very large influence in the $B_s^0\pi^+$ distribution, this contribution is neglected in the current work.

\noindent\emph{$\;$Summary}.\,---\,%
We have investigated the possibility of generating a resonance-like structure $X(5777)$ in the $B_s^0\pi^+$ distribution in reaction $B_c^+\to B_s^0\pi^+\pi^0$. There are several advantages that the proposed rescattering processes may help us to establish a non-resonance interpretation of some $XYZ$ particles, i.e., the TS mechanism. First, the TS kinematical conditions are perfectly fulfilled in those triangle rescattering diagrams. Second, the weak $B\bar{K}$ $(I=1)$ interaction does not support the existence of a narrow dynamically generated resonant or bound state.  Third, all of the relevant couplings in the rescattering diagrams are under good theoretical control, which reduces the model dependence of final results. Further more, the relevant backgrounds in this channel are also expected to be simple. Therefore, if one observes the $X(5777)$ structure in the invariant mass spectrum of  $B_s^0\pi^+$, it is very likely to conclude that this structure originates from the TS and is not a genuine particle. A similar analysis of this paper can be naively extended to the charge conjugate channel $B_c^-\to \bar{B}_s^0\pi^-\pi^0$.  The corresponding experiments should be performed in LHCb. Note, however, a disadvantage for the proposed rescattering processes: there is a neutral pion in the final states. For the LHCb experiments, it is not easy to identify a neutral pion, and thus this poses a severe challenges.

\noindent\emph{$\;$Acknowledgments}.\,---\,%
X. H. Liu is grateful to C. Hanhart for stimulating discussions concerning some of the material presented here. Helpful discussions with L. Y. Dai, C. W. Xiao and W. Wang are also gratefully acknowledged. This work is supported by the DFG and the NSFC through funds provided to the Sino-German CRC 110 ``Symmetries and the Emergence of Structure
in QCD'' (NSFC Grant No. 11261130311).


\begin{thebibliography}{99}

\bibitem{Chen:2016spr} 
H.~X.~Chen, W.~Chen, X.~Liu, Y.~R.~Liu and S.~L.~Zhu,
arXiv:1609.08928 [hep-ph].


\bibitem{Brambilla:2010cs} 
N.~Brambilla {\it et al.},
Eur.\ Phys.\ J.\ C {\bf 71}, 1534 (2011).


\bibitem{Olsen:2014qna} 
S.~L.~Olsen,
Front.\ Phys.\ (Beijing) {\bf 10}, no. 2, 121 (2015)
[arXiv:1411.7738 [hep-ex]].


\bibitem{Brambilla:2014jmp} 
N.~Brambilla {\it et al.},
Eur.\ Phys.\ J.\ C {\bf 74}, no. 10, 2981 (2014).


\bibitem{Chen:2016qju} 
H.~X.~Chen, W.~Chen, X.~Liu and S.~L.~Zhu,
Phys.\ Rept.\  {\bf 639}, 1 (2016).


\bibitem{Esposito:2016noz} 
A.~Esposito, A.~Pilloni and A.~D.~Polosa,
Phys.\ Rept.\  {\bf 668}, 1 (2016).


\bibitem{RMP}
F.-K.~Guo, C.~Hanhart, U.-G.~Mei{\ss}ner, Q.~Wang, Q.~Zhao and B.-S.~Zhou,
commissioned article for Rev. Mod. Phys. (2017).

\bibitem{Chen:2011pv} 
D.~Y.~Chen and X.~Liu,
Phys.\ Rev.\ D {\bf 84}, 094003 (2011).


\bibitem{Bugg:2011jr} 
D.~V.~Bugg,
Europhys.\ Lett.\  {\bf 96}, 11002 (2011).


\bibitem{Swanson:2014tra} 
E.~S.~Swanson,
Phys.\ Rev.\ D {\bf 91}, no. 3, 034009 (2015).


\bibitem{Aitchison:1969tq} 
I.~J.~R.~Aitchison and C.~Kacser,
Phys.\ Rev.\  {\bf 173}, 1700 (1968).


\bibitem{Coleman:1965xm} 
S.~Coleman and R.~E.~Norton,
Nuovo Cim.\  {\bf 38}, 438 (1965).


\bibitem{Bronzan:1964zz} 
J.~B.~Bronzan,
Phys.\ Rev.\  {\bf 134}, B687 (1964).


\bibitem{Schmid:1967ojm} 
C.~Schmid,
Phys.\ Rev.\  {\bf 154}, no. 5, 1363 (1967).


\bibitem{Wu:2011yx} 
J.~J.~Wu, X.~H.~Liu, Q.~Zhao and B.~S.~Zou,
Phys.\ Rev.\ Lett.\  {\bf 108}, 081803 (2012).


\bibitem{Wang:2013cya} 
Q.~Wang, C.~Hanhart and Q.~Zhao,
Phys.\ Rev.\ Lett.\  {\bf 111}, no. 13, 132003 (2013).


\bibitem{Guo:2014iya} 
F.~K.~Guo, C.~Hanhart, Q.~Wang and Q.~Zhao,
Phys.\ Rev.\ D {\bf 91}, no. 5, 051504 (2015).


\bibitem{Ketzer:2015tqa} 
M.~Mikhasenko, B.~Ketzer and A.~Sarantsev,
Phys.\ Rev.\ D {\bf 91}, no. 9, 094015 (2015).


\bibitem{Szczepaniak:2015eza} 
A.~P.~Szczepaniak,
Phys.\ Lett.\ B {\bf 747}, 410 (2015).


\bibitem{Guo:2015umn} 
F.~K.~Guo, U.-G.~Mei{\ss}ner, W.~Wang and Z.~Yang,
Phys.\ Rev.\ D {\bf 92}, no. 7, 071502 (2015).


\bibitem{Liu:2015taa} 
X.~H.~Liu, M.~Oka and Q.~Zhao,
Phys.\ Lett.\ B {\bf 753}, 297 (2016).


\bibitem{Liu:2015fea} 
X.~H.~Liu, Q.~Wang and Q.~Zhao,
Phys.\ Lett.\ B {\bf 757}, 231 (2016).


\bibitem{Achasov:2015uua} 
N.~N.~Achasov, A.~A.~Kozhevnikov and G.~N.~Shestakov,
Phys.\ Rev.\ D {\bf 92}, no. 3, 036003 (2015).


\bibitem{Guo:2016bkl} 
F.~K.~Guo, U.-G.~Mei{\ss}ner, J.~Nieves and Z.~Yang,
Eur.\ Phys.\ J.\ A {\bf 52}, no. 10, 318 (2016).


\bibitem{Roca:2017bvy} 
L.~Roca and E.~Oset,
arXiv:1702.07220 [hep-ph].


\bibitem{Landau:1959fi} 
L.~D.~Landau,
Nucl.\ Phys.\  {\bf 13}, 181 (1959).


\bibitem{Brambilla:2004wf} 
N.~Brambilla {\it et al.} [Quarkonium Working Group],
hep-ph/0412158.


\bibitem{Du:1988ws} 
D.~s.~Du and Z.~Wang,
Phys.\ Rev.\ D {\bf 39}, 1342 (1989).


\bibitem{Sun:2015exa} 
J.~Sun, N.~Wang, Q.~Chang and Y.~Yang,
Adv.\ High Energy Phys.\  {\bf 2015}, 104378 (2015)
[arXiv:1504.01286 [hep-ph]].


\bibitem{Wirbel:1985ji} 
M.~Wirbel, B.~Stech and M.~Bauer,
Z.\ Phys.\ C {\bf 29}, 637 (1985).


\bibitem{Olive:2016xmw} 
C.~Patrignani {\it et al.} [Particle Data Group],
Chin.\ Phys.\ C {\bf 40}, no. 10, 100001 (2016).


\bibitem{Liu:2012zya} 
L.~Liu, K.~Orginos, F.~K.~Guo, C.~Hanhart and U.-G.~Mei{\ss}ner,
Phys.\ Rev.\ D {\bf 87}, no. 1, 014508 (2013).


\bibitem{Mohler:2013rwa} 
D.~Mohler, C.~B.~Lang, L.~Leskovec, S.~Prelovsek and R.~M.~Woloshyn,
Phys.\ Rev.\ Lett.\  {\bf 111}, no. 22, 222001 (2013).


\bibitem{Guo:2006fu} 
F.~K.~Guo, P.~N.~Shen, H.~C.~Chiang, R.~G.~Ping and B.~S.~Zou,
Phys.\ Lett.\ B {\bf 641}, 278 (2006).


\bibitem{Guo:2009ct} 
F.~K.~Guo, C.~Hanhart and U.-G.~Mei{\ss}ner,
Eur.\ Phys.\ J.\ A {\bf 40}, 171 (2009).


\bibitem{Liu:2009uz} 
Y.~R.~Liu, X.~Liu and S.~L.~Zhu,
Phys.\ Rev.\ D {\bf 79}, 094026 (2009).


\bibitem{Kolomeitsev:2003ac} 
E.~E.~Kolomeitsev and M.~F.~M.~Lutz,
Phys.\ Lett.\ B {\bf 582}, 39 (2004).


\bibitem{Altenbuchinger:2013vwa} 
M.~Altenbuchinger, L.-S.~Geng and W.~Weise,
Phys.\ Rev.\ D {\bf 89}, no. 1, 014026 (2014).


\bibitem{Guo:2015dha} 
Z.~H.~Guo, U.-G.~Mei{\ss}ner and D.~L.~Yao,
Phys.\ Rev.\ D {\bf 92}, no. 9, 094008 (2015).


\bibitem{Lu:2016kxm} 
J.~X.~Lu, X.~L.~Ren and L.~S.~Geng,
Eur.\ Phys.\ J.\ C {\bf 77}, no. 2, 94 (2017).


\bibitem{D0:2016mwd} 
V.~M.~Abazov {\it et al.} [D0 Collaboration],
Phys.\ Rev.\ Lett.\  {\bf 117}, no. 2, 022003 (2016).


\bibitem{Albaladejo:2016eps} 
M.~Albaladejo, J.~Nieves, E.~Oset, Z.~F.~Sun and X.~Liu,
Phys.\ Lett.\ B {\bf 757}, 515 (2016).


\bibitem{Aaij:2016iev} 
R.~Aaij {\it et al.} [LHCb Collaboration],
Phys.\ Rev.\ Lett.\  {\bf 117}, no. 15, 152003 (2016)
Addendum: [Phys.\ Rev.\ Lett.\  {\bf 118}, no. 10, 109904 (2017)].


\bibitem{Burns:2016gvy} 
T.~J.~Burns and E.~S.~Swanson,
Phys.\ Lett.\ B {\bf 760}, 627 (2016).


\bibitem{Guo:2016nhb} 
F.~K.~Guo, U.-G.~Mei{\ss}ner and B.~S.~Zou,
Commun.\ Theor.\ Phys.\  {\bf 65}, no. 5, 593 (2016).


\bibitem{Yang:2016sws} 
Z.~Yang, Q.~Wang and U.-G.~Mei{\ss}ner,
Phys.\ Lett.\ B {\bf 767}, 470 (2017).


\bibitem{Oller:2000fj} 
J.~A.~Oller and U.-G.~Mei{\ss}ner,
Phys.\ Lett.\ B {\bf 500}, 263 (2001).


\bibitem{Oller:2000ma} 
J.~A.~Oller, E.~Oset and A.~Ramos,
Prog.\ Part.\ Nucl.\ Phys.\  {\bf 45}, 157 (2000).


\bibitem{Oller:1997ng} 
J.~A.~Oller, E.~Oset and J.~R.~Pelaez,
Phys.\ Rev.\ Lett.\  {\bf 80}, 3452 (1998).


\bibitem{Au:1986vs} 
K.~L.~Au, D.~Morgan and M.~R.~Pennington,
Phys.\ Rev.\ D {\bf 35}, 1633 (1987).


\bibitem{Dai:2012pb} 
L.~Y.~Dai, M.~Shi, G.~Y.~Tang and H.~Q.~Zheng,
Phys.\ Rev.\ D {\bf 92}, no. 1, 014020 (2015).


\bibitem{Klingl:1996by} 
F.~Klingl, N.~Kaiser and W.~Weise,
Z.\ Phys.\ A {\bf 356}, 193 (1996).



\bibitem{Daub:2012mu} 
  J.~T.~Daub, H.~K.~Dreiner, C.~Hanhart, B.~Kubis and U.-G.~Mei{\ss}ner,
  JHEP {\bf 1301}, 179 (2013).

\bibitem{Daub:2015xja} 
J.~T.~Daub, C.~Hanhart and B.~Kubis,
JHEP {\bf 1602}, 009 (2016).



\bibitem{Chen:2015jgl} 
Y.~H.~Chen, J.~T.~Daub, F.~K.~Guo, B.~Kubis, U.~G.~Mei{\ss}ner and B.~S.~Zou,
Phys.\ Rev.\ D {\bf 93}, no. 3, 034030 (2016).

\bibitem{Chen:2016mjn} 
Y.~H.~Chen, M.~Cleven, J.~T.~Daub, F.~K.~Guo, C.~Hanhart, B.~Kubis, U.~G.~Mei{\ss}ner and B.~S.~Zou,
Phys.\ Rev.\ D {\bf 95}, no. 3, 034022 (2017).



\end{thebibliography}
\end{document}